\documentstyle[11pt,aaspp4,epsfig]{article}

\def\ltsima{$\; \buildrel < \over \sim \;$}
\def\simlt{\lower.5ex\hbox{\ltsima}}
\def\gtsima{$\; \buildrel > \over \sim \;$}
\def\simgt{\lower.5ex\hbox{\gtsima}}

\begin{document}
\title{ Studying circumnuclear matter in AGN with X-ray spectroscopy}
\author{Giorgio Matt}
\affil{Dipartimento di Fisica ``E. Amaldi", Universit\`a degli Studi 
``Roma Tre", Via della Vasca Navale 84, I--00146 Roma, Italy}

\begin{abstract}
I discuss the advances in our understanding of the physics and morphology
of the innermost regions of AGNs
which will be possible thanks to the XMM unprecedented sensitivity and its
moderate to high energy resolution.
\end{abstract}

\section{Introduction}

Before entering into details, it may be useful to recall 
the basic picture of the nuclear X--ray emission of 
Seyfert galaxies after the pre-XMM satellites
(e.g. Mushotzky, Done \& Pounds 1993; Fabian 1996; Matt 1998). 
The main component is a power law, possibly cut--offing at energies of
order of a few hundred keV, 
and very likely due to Inverse Compton by 
relativistic electrons of UV/soft X--ray photons, the latter
possibly emitted by the accretion disc (see e.g. Svensson 1996 for a review).
A significant fraction of this  radiation is intercepted and
reprocessed by the accretion disc. The spectrum emerging from the cold matter
is composed by a broad hump peaking at around 30 keV (the Compton reflection
continuum) and a strong iron 6.4 keV fluorescent
line. Kinematic and gravitational effects in the
inner accretion disc are apparent in the line profile. 

In this paper I will assume the unification model for Seyfert
galaxies (Antonucci 1993): type 1 and 2 objects are intrinsically
identical (at least as far as the nuclear properties are concerned),
and surrounded by a $\sim$pc-scale molecular torus. 
If the line--of--sight does not intercept the torus, the nucleus
can be  directly observed and the source is classified as type 1; if
the line--of--sight is blocked by the torus, the source is classified as
type 2. I will come back to unification models in Sec.~2.
It is worth noticing here that the torus may also reprocess part of
the primary radiation,
giving rise to a component similar to that from the accretion disc.
More on this point in Sec.~2.2

The paper is organized as follows: Sec.~2 is devoted to Seyfert 1 galaxies;
both warm absorbers and the reprocessing from the torus are
discussed. In Sec.~3 I will discuss 
Seyfert 2 galaxies, and in particular Compton--thick ones. Sec.~4 will
ve devoted to the study of faint AGNs.

\section{Seyfert 1 Galaxies}

Reprocessing from the accretion disc, and in particular the relativistic
effects on the iron line profile, have been discussed by Fabian in this 
conference. I will therefore concentrate on more distant matter, and
in particular on the ``warm absorber" and on the torus. 

\subsection{Warm absorbers}

\subsubsection{Generalities}

``Warm absorbers", i.e. (photo)ionized matter on the line--of--sight, 
were introduced by Halpern (1984) to account for the spectral variability 
observed in the quasar MR 2251--178. A spectacular and more
direct confirmation of the existence of circumnuclear ionized matter
came from the ROSAT/PSPC observation of MCG--6-30-15,
the Seyfert 1 galaxy which is so famous
for its relativistic line (Tanaka et al. 1995; Fabian, this conference):
oxygen absorption edges were clearly detected (Nandra \& Pounds 1992). 
The next step forward came with ASCA, which for the first time has been able,
 thanks to the improved energy resolution provided by the CCDs,  to
resolve the O~{\sc vii} and O~{\sc viii} absorption edges (Fabian et al.
1994). Systematic studies of warm absorbers in Seyfert 1 galaxies with ASCA
have permitted to establish their commonness, being detected in at least
half of the sources (Reynolds 1997; George et al. 1998). I will come back
to this point in Sec.~2.1.2.

What are the signatures of warm absorbers? Detailed models 
have been produced by several authors (e.g. Netzer 1993; Reynolds \&
Fabian 1995; Krolik \& Kriss 1995; Nicastro et al. 1998a; Nicastro, Fiore
\& Matt 1998; Porquet et al. 1998). Since the early days, it was clear
that the best observable features are the photoelectric absorption edges, and
up to now they are in fact 
the only features unambiguously detected. Netzer (1993)
argued that emission lines are also to be expected, their intensity 
depending on the covering factor of the ionized matter. Matt (1994) and
Krolik \& Kriss (1995) pointed out the potential importance of absorption
lines. While absorption edges are observable with moderate energy resolution
but sensitive instruments (which is the reason why they have been so
well studied by ASCA, and detected even by proportional counters),
both emission and absorption lines require high energy resolution 
detectors, like the gratings. I will show below how  their
diagnostic capabilities can be exploited by the XMM/RGS. 

Usually, warm absorbers are described by single zone, photoionization 
equilibrium models. In this situation, simple variability patterns are
expected: increasing ionization of the matter when the source brightens.
As usual, nature is more complex. The best example is again MCG--6-30-15.
During the long--look ASCA observation (Otani et al. 1996)
the source varied with large amplitude on short time scales. The depth
of the  O~{\sc vii} edge, on the contrary, stayed
 almost constant, while that of the O~{\sc viii} edge was roughly 
anticorrelated
with the flux. While the last behaviour could be easily explained (higher
the flux, higher the fraction of fully stripped iron at the expense of
H--like one), the behaviour of the  O~{\sc vii} edge was completely at
odds with any simple model. The author's suggestion is that there are
actually two different regions were warm absorption occurs. The first one 
would be very close to the nucleus and highly ionized, and 
responsible for the bulk 
of O~{\sc viii} absorption. The second region would be more distant,
less ionized and rather tenuous, so that the recombination time scale would be
longer than the variability time scale, i.e. the matter would 
not be in photoionization
equilibrium. Lack of photoionization equilibrium was invoked by Orr
et al. (1997) also for the first zone to account for the fact that 
during the BeppoSAX observation also the O~{\sc viii} edge versus flux
anticorrelation disappeared. Warm absorbers out of equilibrium have
been studied in some detail by Nicastro et al. (1998a), who show that
different ions respond on different time scales to variations in the
ionizing continuum, so confusing attempts to derive the properties
of the warm absorbers. 

\begin{figure}[t]
\centerline{\epsfig{file=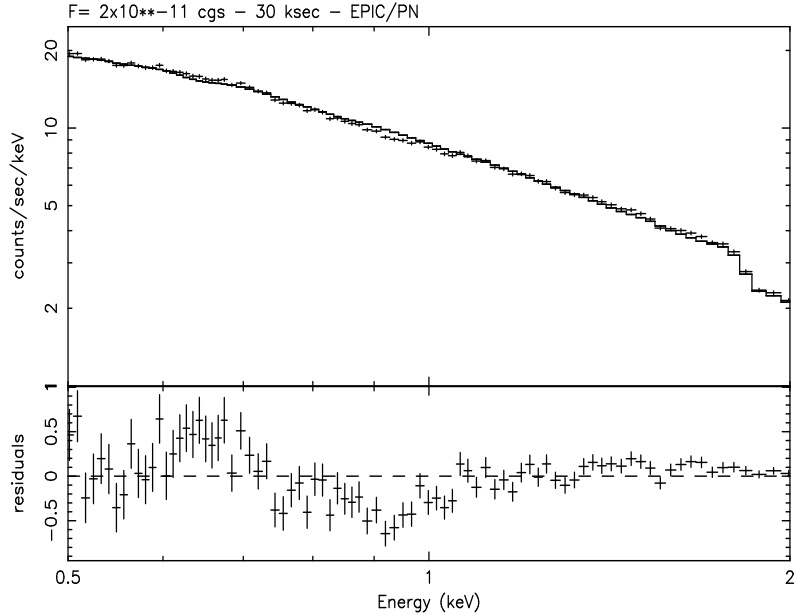,width=10cm,height=12cm,angle=-90}}
  \caption{EPIC/PN simulated spectrum of a warm absorber source (see text). 
The fit is performed with a simple power law. Oxygen edges are apparent
in the residuals.}
\label{edge}
\end{figure}

\subsubsection{Are warm absorbers ubiquitous?}

Reynolds (1997) and George et al. (1998) studied a large sample of
Seyfert 1 galaxies observed by ASCA. Evidence for ionized absorption
(i.e. absorption edges, especially of oxygen ones)
was found in at least half of the sources, indicating that warm 
absorbers are rather common. Common or ubiquitous? There are at least
three possibilities to explain the lack of evidence for ionized absorption 
in the other half of the sample:

\begin{itemize}

\item The optical depth of the absorption edges is too small to be detectable
(as an extreme case, it may be zero, i.e. ionized matter would be absent 
altogether). Typical upper limits for the O~{\sc vii} edge are around
0.1 for bright sources. 
The much larger sensitivity of XMM/EPIC with respect to the ASCA/SIS
will permit to probe much lower column densities. As an example, in 
Fig.~\ref{edge} we show the spectrum and
residuals when a simple power law is fitted to
a simulated 30 ksec EPIC/PN spectrum including O~{\sc vii} and O~{\sc viii} 
absorption edges with $\tau$=0.07. The 2--10 flux of the source is 
2$\times10^{-11}$ erg cm$^{-2}$ s$^{-1}$, i.e. bright but not extreme. 
A 2$\times10^{20}$ cm$^{-2}$ column density has been included 
to account for Galactic absorption. 
The oxygen edges are apparent in the residuals. The
optical depths of the edges are recovered with $\sim$20\% accuracy.
 
\item A different explanation 
may be that in these sources the matter is so highly ionized
to significantly reduce the optical depths of  O~{\sc vii} and O~{\sc viii}
edges, i.e. the best signatures of warm absorbers with ASCA. 
Small, residual  O~{\sc viii} absorption edges and/or iron edges should be
then searched for. Again, high sensitivity is required, and XMM/EPIC
is the most suited instrument in the near future.

\begin{figure}[t]
\centerline{\epsfig{file=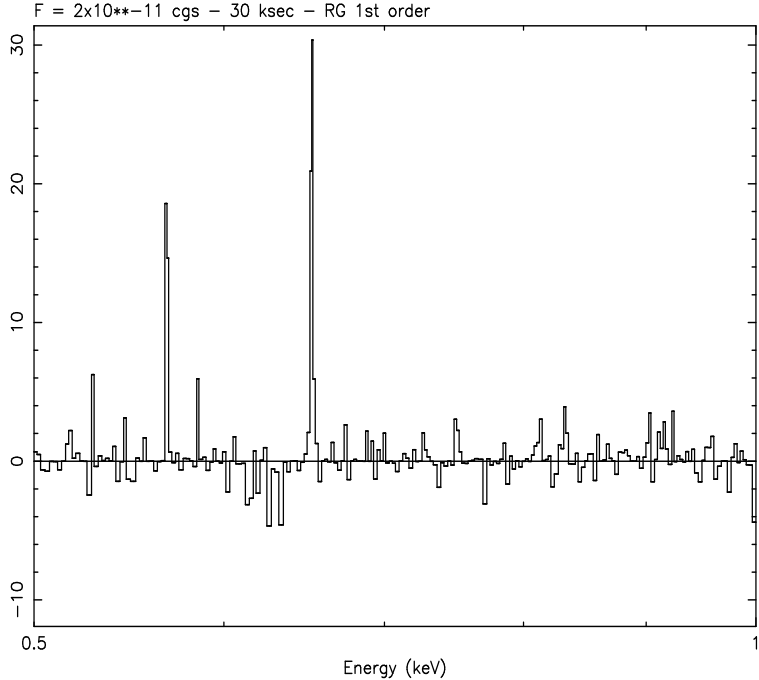,width=10cm,height=12cm,angle=-90}}
  \caption{RGS simulated spectrum of a power law plus He-- and H--lile 
oxygen emission lines with EW=3 eV each. The plot show the residuals after
fitting the spectrum with a simple power law.}
\label{lines}
\end{figure}

\item Finally, it may be possible that the covering factor 
of the warm absorbers is less than 1.
In this case, warm absorbers are detected only in those sources in which 
they happen to be on the line of sight. In the other cases, however,
warm absorbers may be indirectly observed through emission lines, which are
expected to be produced in the warm material (Netzer 1993). 
Unfortunately, line equivalent widths (EWs) are expected to be very
small, hopelessly so for moderate energy resolution detectors like
CCDs. With the gratings, however, it is a different story. Lines should be
narrow enough (i.e. $\Delta E\sim$ a few eV for oxygen lines
emitted in region with dispersion velocities typical of the Broad Line
Regions) to be a perfect target for the XMM/RGS. This is demonstrated in 
Fig.~\ref{lines}, where a power law spectrum plus
O~{\sc vii} and O~{\sc viii} narrow (i.e width set to zero for simplicity)
emission lines is shown. Parameters of the source are as in Fig.~1.
The simulation is for
the first order RGS only; the exposure time is 30 ksec. The EWs of the lines
are 3 eV each, a reasonable value for typical warm absorbers (Netzer 1996). 
The detection of the lines is highly significant: warm absorbers may 
be successfully searched for even if we are not observing absorption, after all! 

\end{itemize}

\subsubsection{Warm absorbers in Narrow Line Seyfert 1 galaxies}

Narrow Line Seyfert 1 galaxies (NLS1s: e.g. Boller, Brandt \& Fink 1996; see
Brandt 1998 for a recent review) are a subclass of Seyfert 1s 
with Broad Lines narrower than usual (but still broader
than the Narrow Lines). Apart from this rather confusing (at least
semantically) optical properties, they differ in X--rays from classical
Seyfert 1s because of a much steeper spectrum in soft X--rays
(and somewhat steeper in hard X--rays, too) as well as a larger
variability.

\begin{figure}[t]
\begin{minipage}{75mm}
\epsfig{file=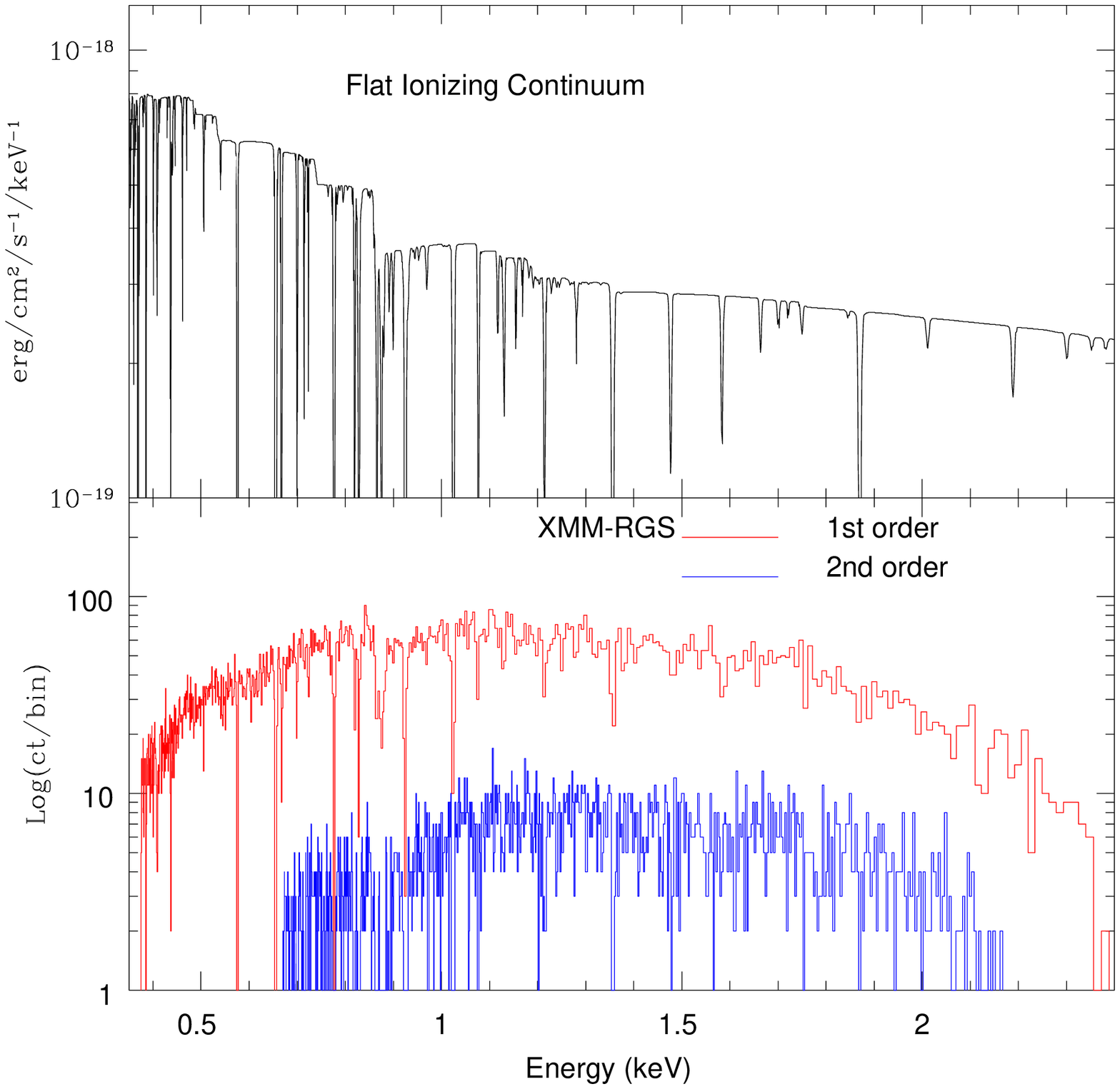,width=75mm}
\end{minipage}
\hspace*{1cm}
\begin{minipage}{75mm}
\epsfig{file=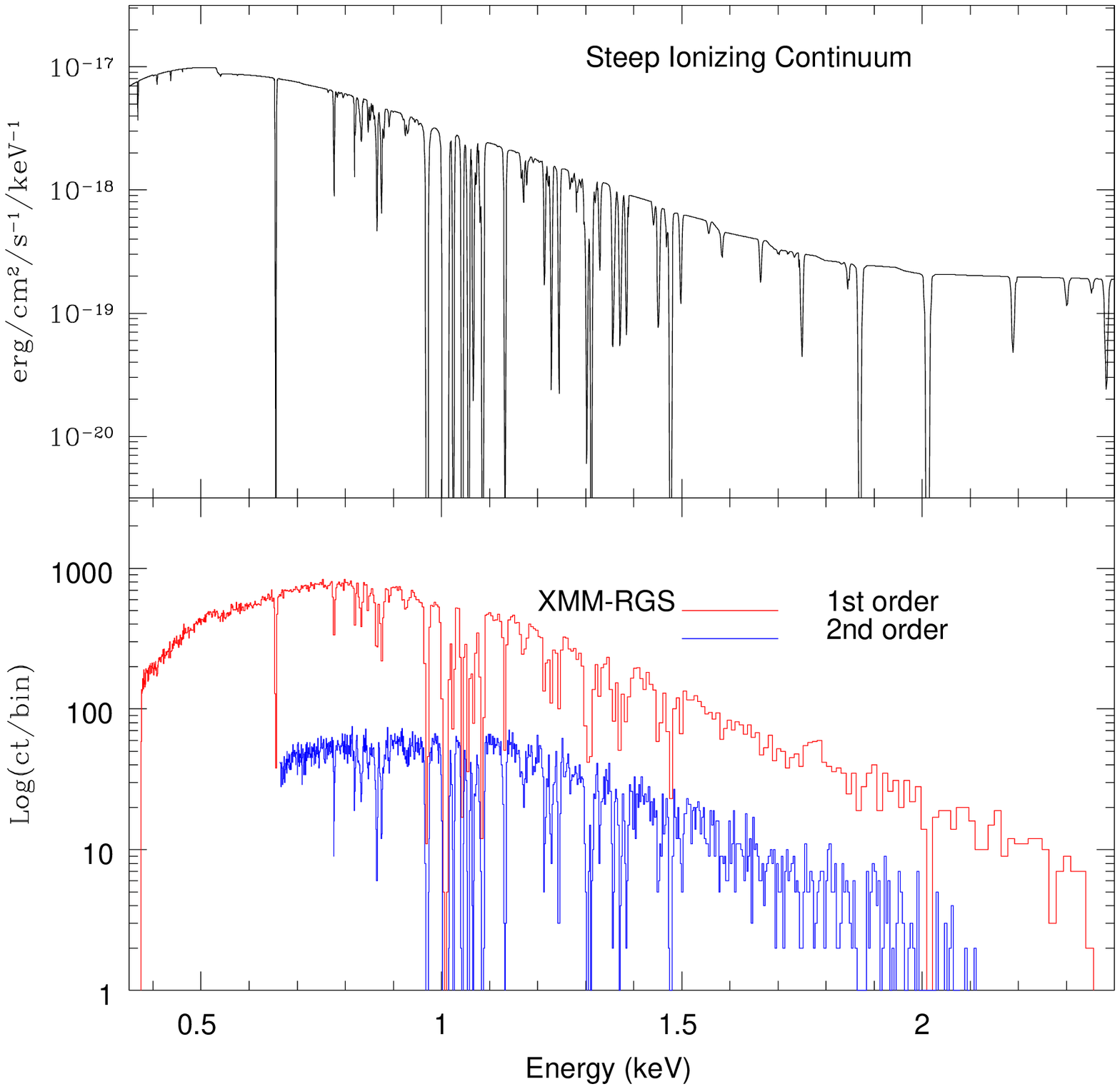,width=75mm}
\end{minipage}
\caption{Theoretical (upper panel) and RGS simulated (lower panel) spectra
of warm absorbers from a flat spectrum source (i.e. a classical Seyfert 1,
left panel) and a steep spectrum source (i.e. a NLS1, right panel). See
Nicastro, Fiore \& Matt 1998, and this conference.}
\label{reso}
\end{figure}

Recently, absorption--like features around 1 keV have been
detected in the ASCA spectra of some NLS1 
(Leighly et al. 1997; Fiore et al. 1998a; Nicastro et al. 1998b). The 
most obvious explanation would be in terms of neon absorption edges, but it  
clashes with the lack of any oxygen edge, and
must be abandoned unless resorting to unusually high Ne/O abundances
(Komossa \& Fink 1998). Alternative explanations include
blueshifted ($z\sim$0.2-0.3) oxygen edges (Leighly et al. 1997) or
line emission from Ne and/or Fe L (Fiore et al. 1998a). In my opinion,
however, the most promising
explanation, and one which can be easily tested with XMM/RGS, 
involves resonant absorption lines (Nicastro, Fiore \& Matt
1998 and this conference). This explanation answers also to the
question of why these features are observed only in NLS1s. The key is
in the much steeper spectrum. Flat spectra, like those typical of 
classical Seyfert 1s, produce an ionization structure in which, if iron
is mildly ionized, oxygen is highly but not completely ionized 
(Fig.~\ref{reso}, left panel). Many absorption lines are present, but
edges remains the best observable feature for moderate resolution 
detectors. For typical NLS1 spectra (i.e. very steep in soft
X--rays, flatter above 1--2 keV), on the contrary, when iron is
mildly ionized oxygen is almost completely stripped: absorption edges
are no longer important, and the main absorption feature is the iron
L forest (Fig.~\ref{reso}, right panel) which, when
convolved with the ASCA/SIS response matrix, may
be observable as a broad absorption feature around 1 keV (Nicastro, Fiore 
\& Matt 1998). In the lower panels of Fig.~\ref{reso} simulations with 
both the first and second order XMM/RGS are shown: the resonant absorption
model will be thoroughly tested by XMM. 

\subsection{Searching for the torus in Seyfert 1s}

In recent years, it has became evident 
that some intrinsic differences between the average properties of
Seyfert 1s and 2s do exist, contrary to the strict version of 
unification models:  enhanced star formation
in Seyfert 2 galaxies (Maiolino et al. 1997); different
morphologies between galaxies hosting type 1 and 2 nuclei, those hosting
type 2 being on average more irregular (Maiolino et al. 1997, Malkan et al.
1998); a greater dust content in Seyfert 2s
(Malkan et al. 1998). X--ray observations of optically selected samples
have demonstrated that the nuclei of Seyfert 2 galaxies are usually
 strong X--ray emitter (e.g. Maiolino et al. 1998), which implies that
the differences most likely are in the environment rather than in the nucleus.

\begin{figure}[t]
\centerline{\epsfig{file=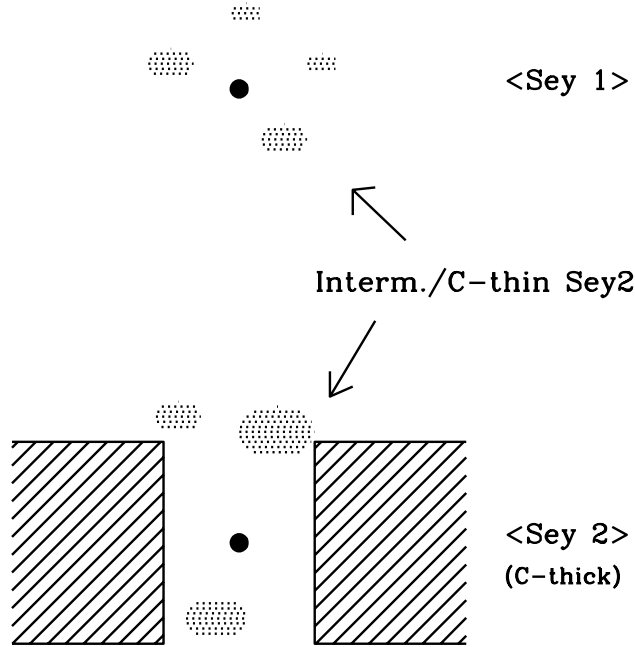,width=11cm,height=11cm}}
\caption{Possible modification for the unification model: 
the average Seyfert 1 does not possess a torus (but a fraction Seyfert 1s with
a torus are also expected), which is typical of 
Compton--thick Seyfert 2s. Intermediate Seyfert and Compton--thin
Seyfert 2s are those objects observed through optically thin, to Compton
scattering, dust lanes.}
\label{um}
\end{figure}

A possible modification of the unification model may be as illustrated in 
Fig.~\ref{um} (see also Matt 1998).  
In this scenario, typical Seyfert 1s would not possess
a (Compton--thick) torus, even if some objects {\it with} the torus are
expected. To test the model, it is then important to search for
torii in Seyfert 1s. When the torus (as any neutral matter) is illuminated
by X--rays, the reprocessing of photons gives rise to the 
iron fluorescent line at
6.4 keV and the Compton reflection continuum (e.g. Lightman \& White 1988;
George \& Fabian 1991; Matt, Perola \& Piro 1991; Ghisellini, Haardt 
\& Matt 1994). The problem is that the inner accretion disc produces
similar features. Torus and disc reprocessing 
may be distinguished in two ways:

\begin{itemize}

\item By searching for delayed variability in the reprocessing component.
While any contribution from the disc should respond almost simultaneously to
the primary flux (say, within a day or so), the reprocessing from the
torus should vary on time scales of months or even years.  A monitoring
campaign would then be required. Of course, the most dramatic the variation
in the primary flux, the easier the detection of any delayed reprocessing
component. The ideal case would be the switching off of the central
engine. This 
actually occurred in NGC~4051 (Guainazzi et al. 1998a), which has been observed
by BeppoSAX on May 1998. Fig.~\ref{4051} shows the observed spectrum
(right panel), and the comparison of the best fit model
with the 1994 ASCA observation (right
panel). In the BeppoSAX observation the nucleus is completely
invisible, and the spectrum is well fitted by a pure reflection 
component. This is probably the best evidence for substantial amount
of circumnuclear cold matter in a Seyfert 1 galaxy. 

\begin{figure}[t]
\begin{minipage}{70mm}
\epsfig{file=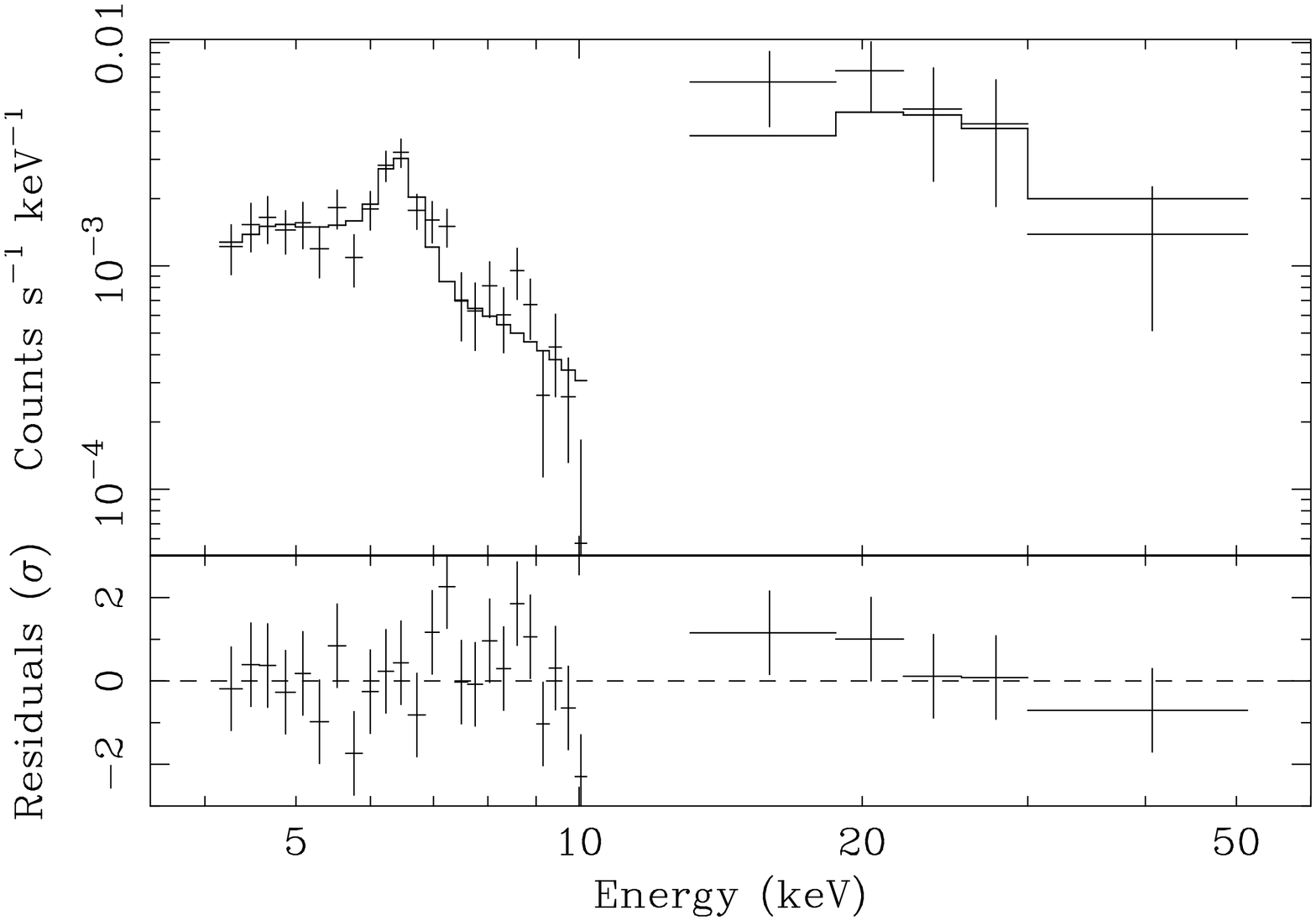,width=65mm,height=70mm,angle=-90}
\end{minipage}
\hspace*{1cm}
\begin{minipage}{70mm}
\epsfig{file=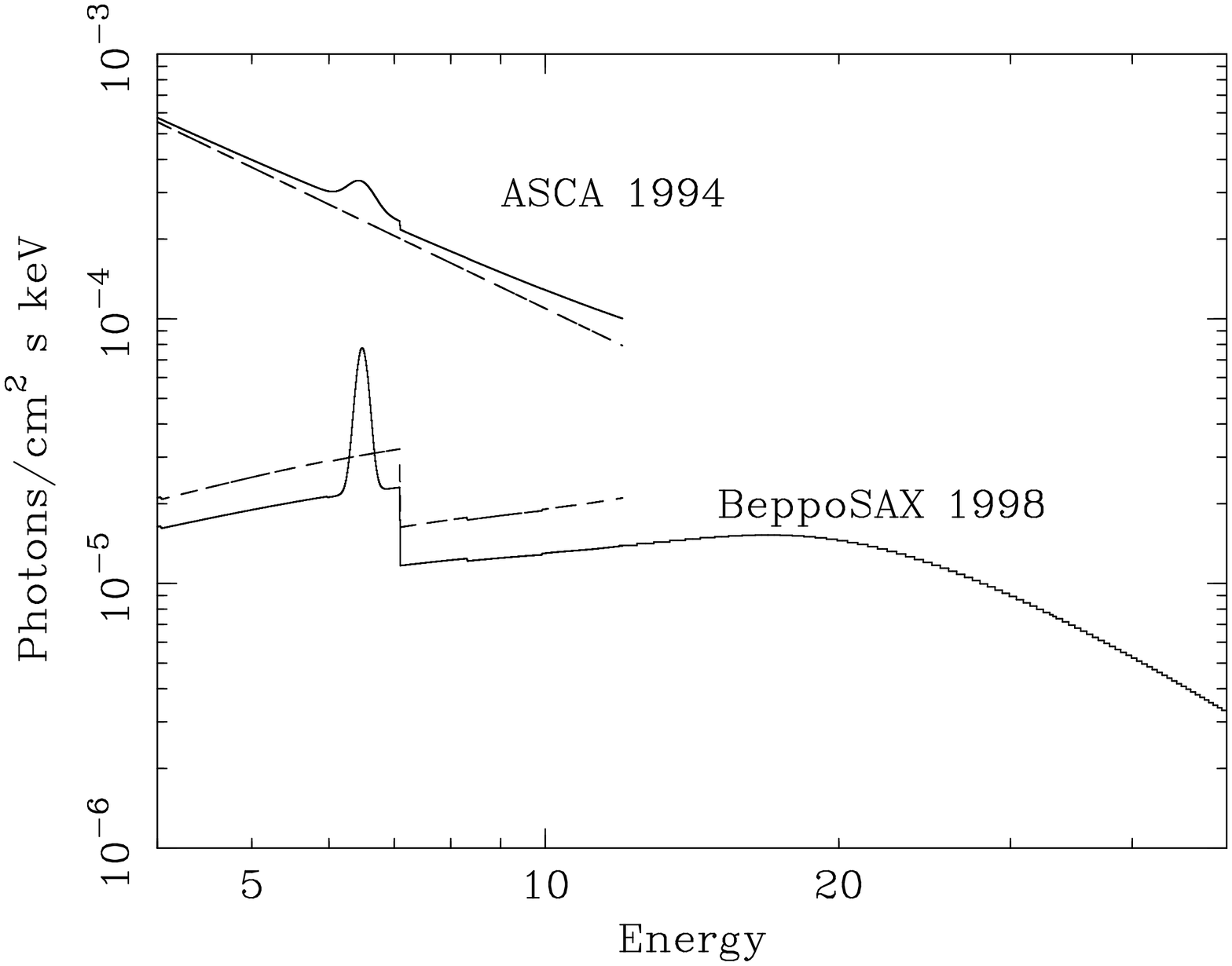,width=65mm,height=70mm,angle=-90}
\end{minipage}
\caption{Left panel: May 1998 BeppoSAX spectrum of NGC~4051. Right panel:
bets fit model, compared with the 1994 ASCA one. From
Guainazzi et al. (1998a)}
\label{4051}
\end{figure}

\item As the NGC~4051 behaviour has been unique so far, one has to consider 
also the other method to distinguish between disc and torus
reprocessing: the profile of the iron line. Lines from the inner
accretion discs are significantly broadened and skewed by relativistic
effects (Fabian et al. 1989; Matt et al. 1992), while those from the
torus should be narrow. While in many sources the presence of a broad
line is out of question (Tanaka et al. 1995; Nandra et al. 1997; Fabian,
this conference), the presence of a further narrow component is not
so well established.
Nandra et al. (1997) found that a contribution to the line
from the torus is usually not required but also not ruled out by the
data. Guainazzi et al. (1998b) found an upper limit of 40 eV to 
any narrow contribution in the BeppoSAX data on MCG--6-30-15, somewhat lower
than expected from the torus (Ghisellini, Haardt \& Matt 1994). Of course, to
search for narrow line components high energy
resolution, high sensitivity detectors are needed. Even if ASTRO-E
seems to provide the best compromise between resolution and effective
area at the iron line energy, 
the large sensitivity of XMM/EPIC coupled with its moderate
energy resolution will be valuable in this respect. 

\end{itemize}

\section{Seyfert 2 galaxies}

X--ray spectroscopy has 
already played a vital role in the study of Seyfert 2 galaxies,
and promises to be very fruitful also in the future.
This is particularly true for the subclass of Compton--thick Seyfert 2s, i.e.
sources for which the column density of the line of sight absorber (from now
on identified with the torus) 
is so large ($\simgt10^{24}$ cm$^{-2}$) to be optically thick to 
Compton scattering. In this case, the nucleus is completely obscured in 
soft and medium X--rays (and, if N${\rm H}\simgt10^{25}$ cm$^{-2}$, even in hard
X--rays), and in particular at the iron line energy. This permits
to study components which would otherwise be diluted to
invisibility by the nuclear radiation. 

The best studied case so far is NGC~1068. The X--ray spectrum is rather complex
(Marshall et al. 1993; Ueno et al. 1994, Iwasawa et al. 1997; Matt et al.
1997; Guainazzi et al. 1999). Below $\sim$4 keV, a thermal--like component,
probably associated with the extended starburst emission observed by 
ROSAT/HRI (Wilson et al. 1992), dominates. A prominent  soft X--ray emission
is often found in Seyfert 2s; it is generally not clear whether this emission is
associated with starburst emission or is due to scattering of the nuclear
radiation from warm, photoionized matter. Soft X--ray spectroscopy like that provided
by XMM/RGS will be valuable in this respect, as the line spectrum in the
two cases should be quite different. 

\begin{figure}[t]
\begin{minipage}{70mm}
\epsfig{file=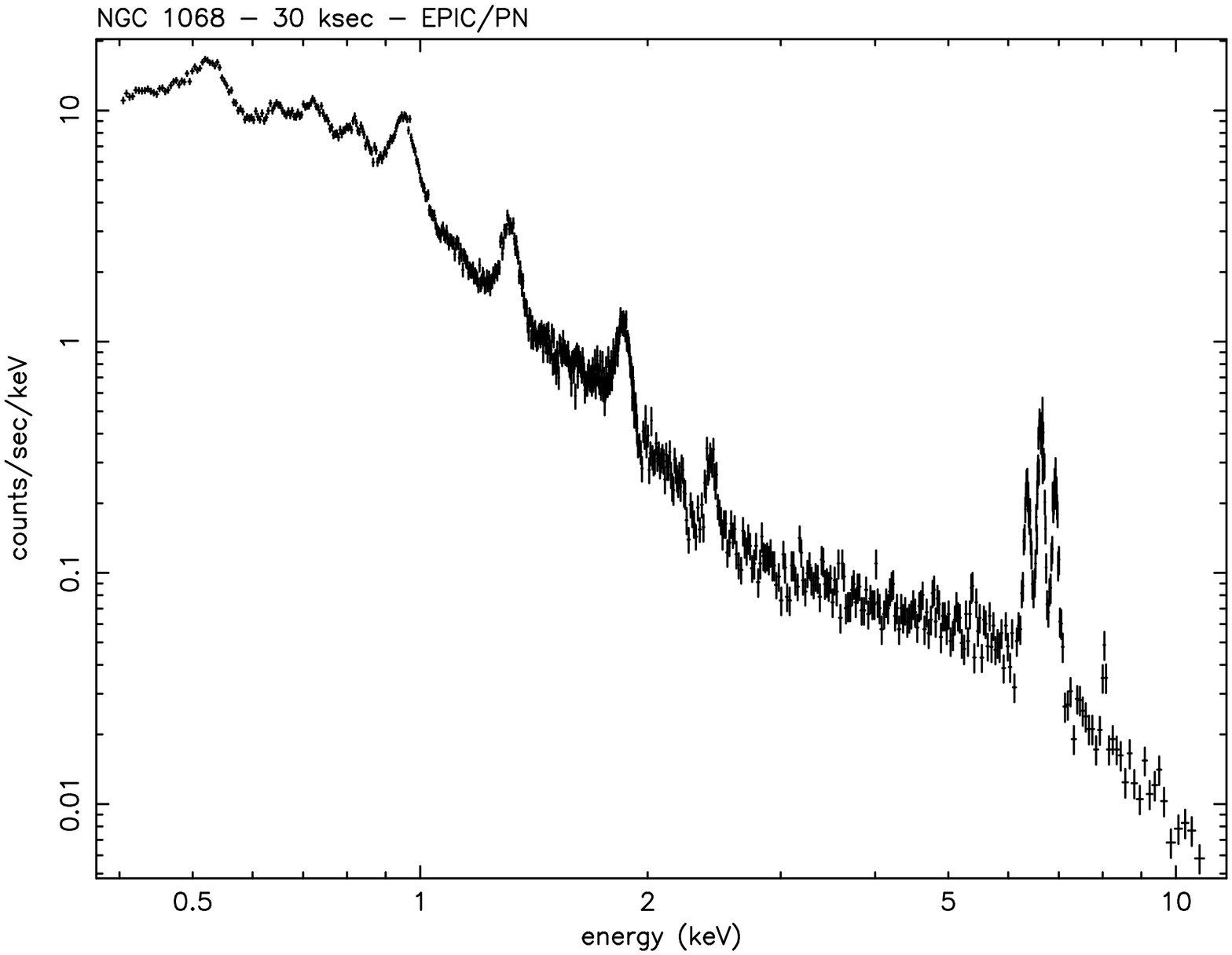,width=65mm,height=70mm,angle=-90}
\end{minipage}
\hspace*{1cm}
\begin{minipage}{70mm}
\epsfig{file=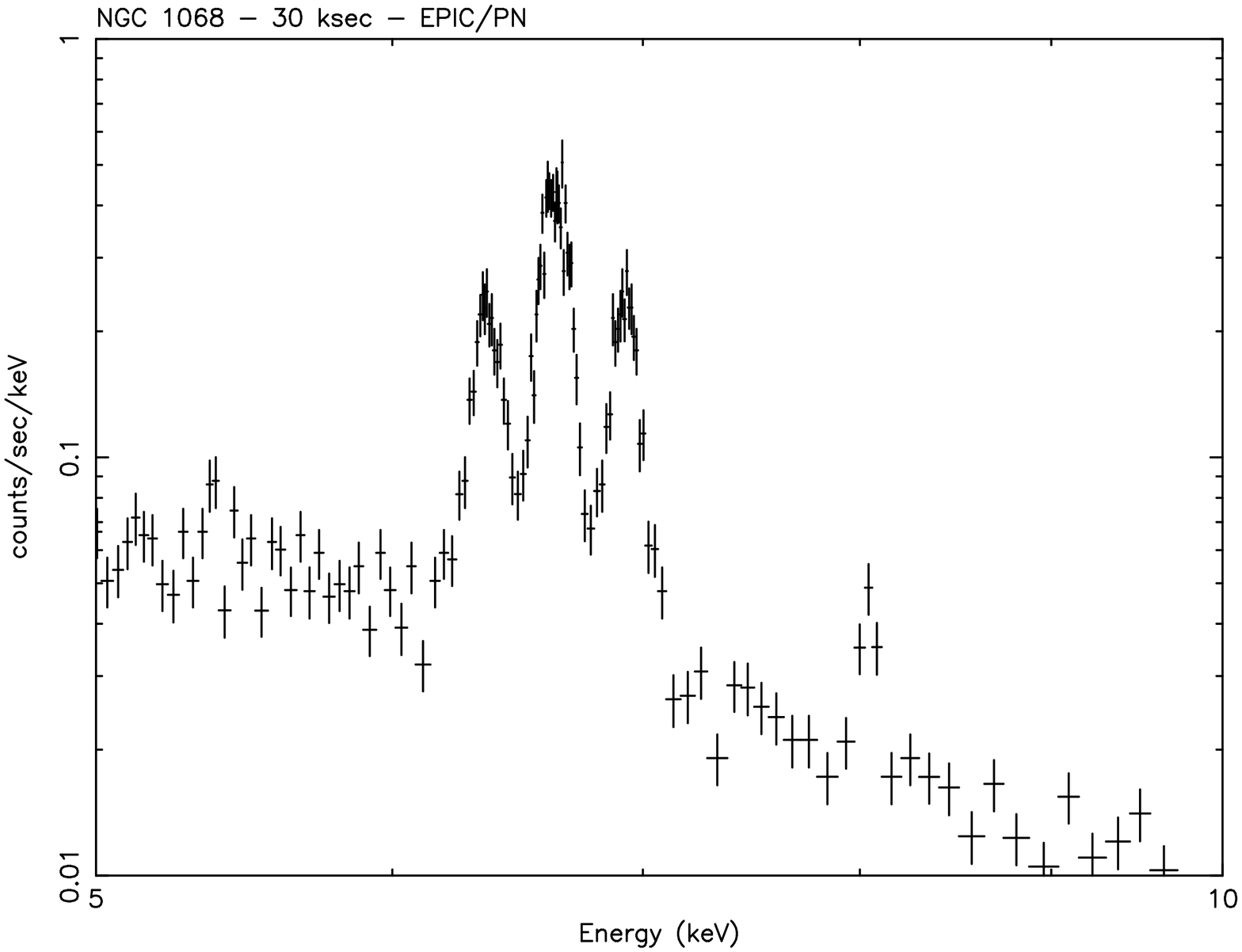,width=65mm,height=70mm,angle=-90}
\end{minipage}
\caption{EPIC/PN simulation of the BeppoSAX best fit model for NGC~1068
(left panel). I the right panel, a blow--up of the iron line complex is 
shown.}
\label{1068}
\end{figure}

Above 4 keV, the spectrum of NGC~1068 is dominated by two components,
both of them due to reflection of the nuclear radiation by circumnuclear 
matter. The first is due to reflection from cold matter, and is characterized
by a Compton reflection continuum plus an intense iron fluorescent line
at 6.4 keV. The second is due to reflection from ionized matter, and is
 characterized by a power law continuum (just mirroring the primary
continuum) and by several lines, most notably the He-- and H--like
iron recombination/resonant scattering lines. From the lines, 
information on the physical and geometrical properties of
the reflectors can be derived (e.g. Matt, Brandt \& Fabian 1996).
A 30 ksec EPIC/PN simulation
of the BeppoSAX best fit spectrum of  NGC~1068 is shown in Fig~\ref{1068}
(left panel, while the blow--up on the iron line complex is shown in
the right panel) to illustrate the quality of the spectrum which can be
obtained with XMM.

\section{Faint AGNs}

\begin{figure}[t]
\begin{minipage}{70mm}
\epsfig{file=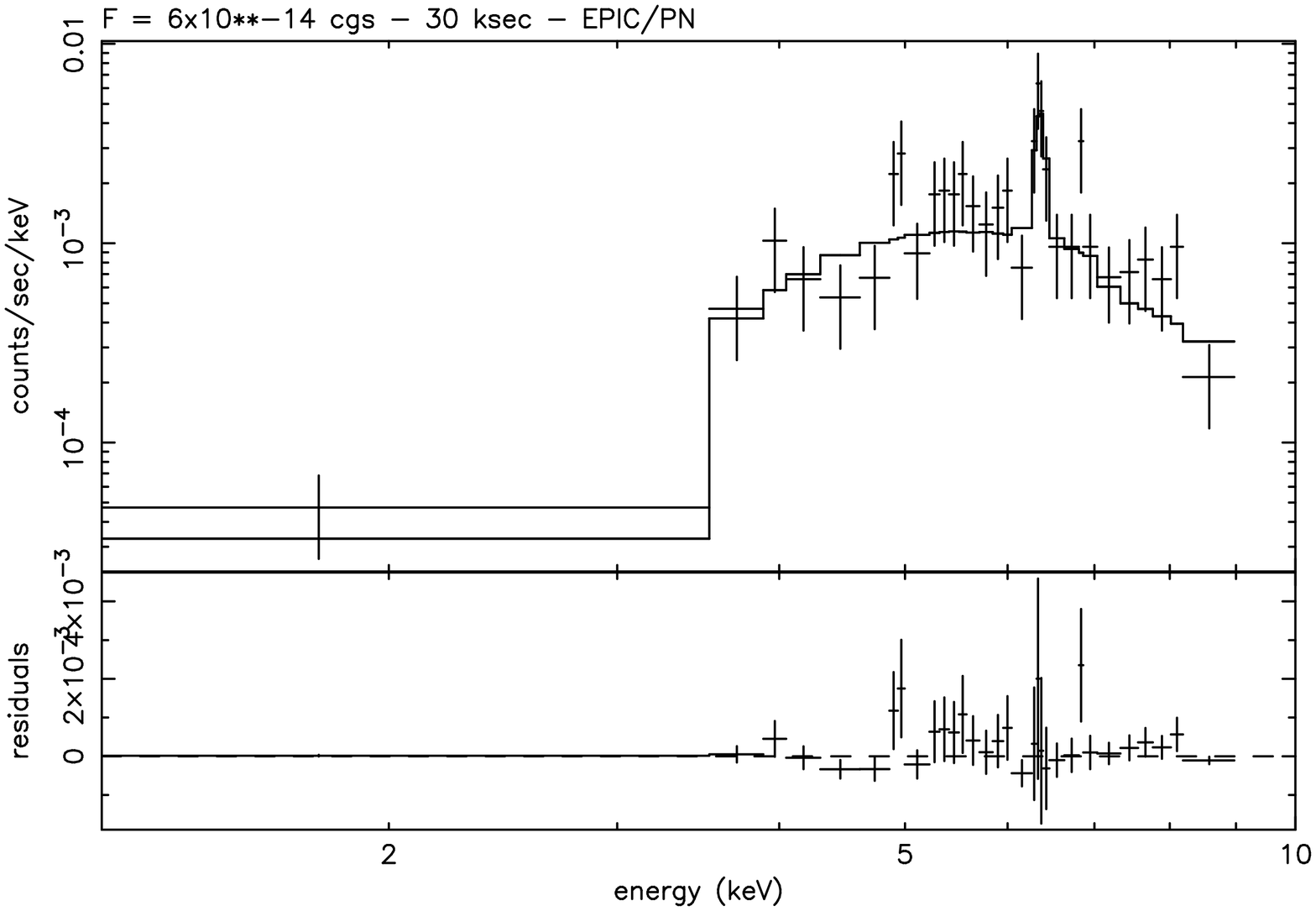,width=65mm,height=70mm,angle=-90}
\end{minipage}
\hspace*{1cm}
\begin{minipage}{70mm}
\epsfig{file=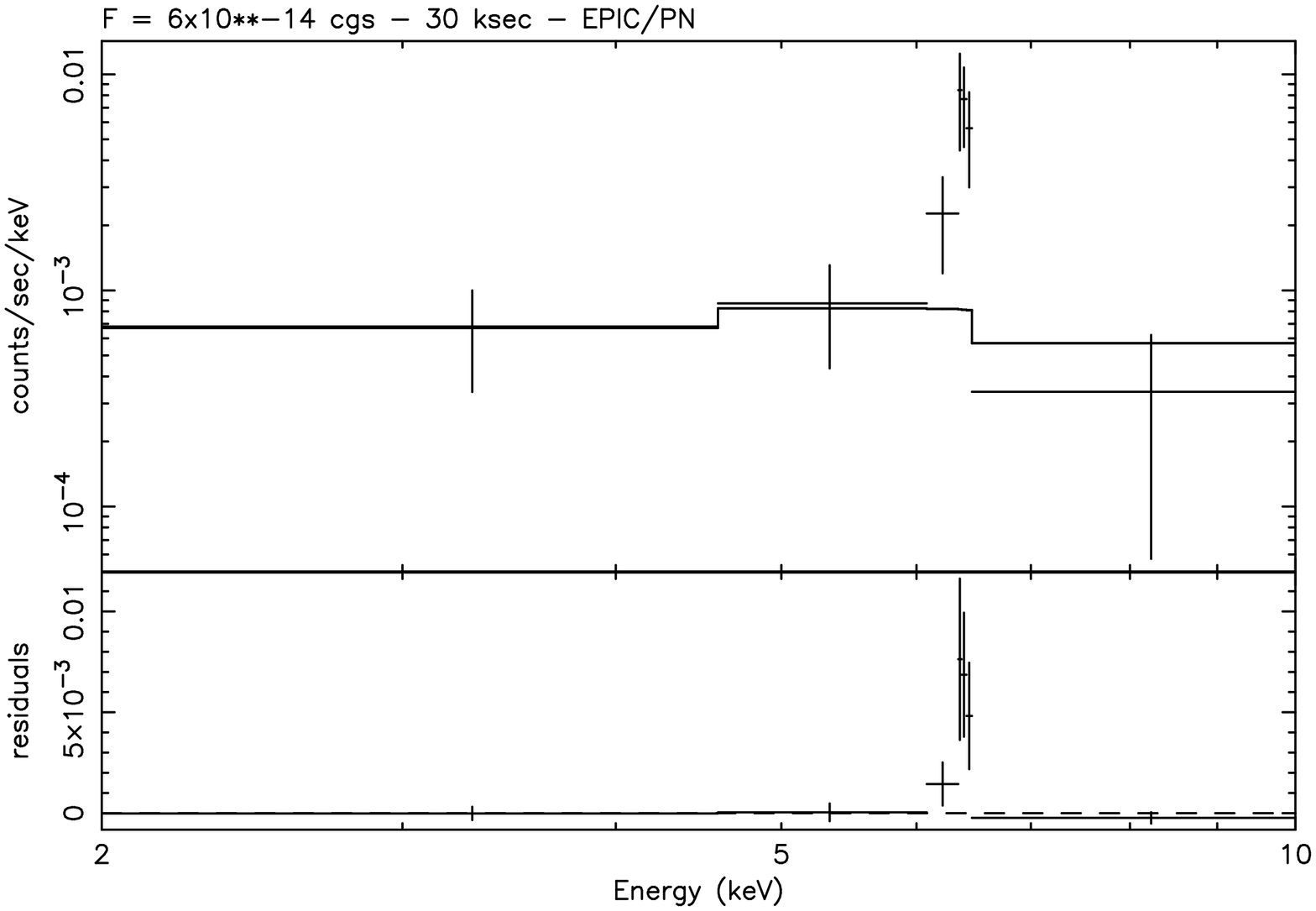,width=65mm,height=70mm,angle=-90}
\end{minipage}
\caption{EPIC/PN simulation of a Compton--thin (left panel) and a
compton--thick (right panel) source. The chosen flux corresponds to
the flux limit of the BeppoSAX/HELLAS survey. For the Compton--thick
source, the fit does not include the iron line.}
\label{faint}
\end{figure}

ASCA (Ueda et al. 1998; Akiyama et al., this conference) 
and BeppoSAX (Fiore et al. 1998b, and this conference) 
have started resolving the hard (i.e above $\sim$2 keV) X--ray Background
(XRB) in discrete sources, many of them turning out to be
AGNs. According to popular synthesis
models of the XRB (e.g. Comastri et al. 1995 and references therein), most
of the AGNs which produce the XRB should be absorbed, i.e. type 2
objects. The fraction of absorbed AGN and their N$_{\rm H}$ distribution down
to very low fluxes can be studied by XMM/EPIC thanks to its unprecedented
sensitivity. In Fig.~\ref{faint} simulations of a Compton--thin
(N$_{\rm H}=3\times10^{23}$ cm$^{-2}$; left panel) and a Compton--thick
(right panel) sources are shown. In both cases, the observed flux is set to
6$\times10^{-14}$ erg cm$^{-2}$ s$^{-1}$, i.e. the flux limit of the
deepest hard X--ray survey so far, i.e. the BeppoSAX HELLAS survey
(Fiore et al. 1998b and this conference). The two kind of sources are
easily distinguishable each other, and the column density, in the first case, 
is recovered with a 50\% precision. It is worth noting that, in the 
Compton--thick case, the iron line energy is recovered with an error of
only 0.1 keV, and therefore the redshift may be determined with great
accuracy. No optical follow--ups will be needed, neither 
for the identification nor for the distance!

\section*{Acknowledgements}

I thank all my collaborators, and in particular Fabrizio Nicastro,
for fruitful discussions.

\end{document}